\documentclass[%
 reprint,
 amsmath,amssymb,
 aps,
]{revtex4-1}

\usepackage{graphicx}
\usepackage{dcolumn}
\usepackage{bm}
\usepackage{float}
\usepackage{color}

\newcommand{\dd}{\mathrm{d}}
\newcommand{\rev}{}

\begin{document}

\preprint{APS/123-QED}

\title{Continuous rotation of achiral nematic liquid crystal droplets driven by heat flux }

\author{Jordi Ign\'es-Mullol$^{a,b}$}
\email{jignes@ub.edu}
\author{Guilhem Poy$^a$}
\author{Patrick Oswald$^a$}
\email{patrick.oswald@ens-lyon.fr}

 \affiliation{%
$^a$Univ Lyon, ENS de Lyon, Univ Claude Bernard, CNRS, Laboratoire de Physique, F-69342 Lyon, France\\
$^b$Universitat de Barcelona, Institute of Nanoscience and Nanotechnology (IN2UB) and Departament de Qu\'{\i}mica F\'{\i}sica, Mart\'i i
Franqu\`es 1, 08028, Barcelona, Spain
 }%

\date{\today}

\begin{abstract}
Suspended droplets of cholesteric (chiral nematic) liquid crystals spontaneously rotate in the presence of a heat flux due to a temperature gradient, a phenomenon known as Lehmann effect. So far, it is not clear whether this effect is due to the chirality of the phase and the molecules or  only to the chirality of the director field.  Here, we report the continuous rotation in a temperature gradient of nematic droplets of a lyotropic chromonic liquid crystal featuring a twisted bipolar configuration. The achiral nature of the molecular components leads to a random handedness of the spontaneous twist, resulting in the coexistence of droplets rotating in the two senses, with speeds proportional to  the temperature gradient and inversely proportional to the droplet radius. This result shows that a macroscopic twist of the director field is sufficient to induce a rotation of the droplets, and that the  phase and the molecules do not need to be chiral. This suggests that one can also explain the Lehmann rotation in cholesteric liquid crystals without introducing the Leslie thermomechanical coupling -- only present in chiral mesophases. An explanation based on the Akopyan and Zeldovich theory of thermomechanical effects in nematics is proposed and discussed.

\begin{description}
\item[PACS numbers]
61.30.St, 65.40.De, 05.70.Ln
\end{description}
\end{abstract}

\maketitle


{\rev Molecular rotors and motors \cite{michl}, such as the proteins that drive active self-organization in the cell cytoskeleton, are nanoscale entities that transform energy from a surrounding source into directed motion. Among the incipient man-made attempts at replicating such behavior, a remarkable example is endowed by chiral-nematic, or cholesteric, liquid crystals (CLC), whose director field can be set into rotation by gradients of scalar fields \cite{degennes}. Recent experimental realizations include chemo-mechanical coupling both in Langmuir monolayers \cite{tabe2,zywo} and in thin films of chiral molecules \cite{seki}, or electro-mechanical coupling in different bulk and disperse systems \cite{gil1,madhusudana,padmini}, although the interpretation of the latter type of experiments often remains unclear \cite{gil2,baudry,kramer,tarasov,dequidt}. Perhaps the best known among these intriguing non-equilibrium thermodynamic phenomena is the thermomechanical coupling, known as Lehmann effect, described in 1900 by Lehmann \cite{lehmann}, who reported the continuous rotation of the internal texture of CLC droplets in equilibrium with the isotropic phase, when they were subjected to a temperature gradient. In spite of the apparent simplicity of these observations, they remained experimentally elusive until recently \cite{oswald1, oswald2, tabe, sano}.

From a theoretical perspective, Leslie's generalization of nematohydrodynamics to CLC \cite{leslie1,leslie2} demonstrated that the absence of mirror and inversion symmetries enable the induction of a torque on the CLC director by means of a thermal gradient. The assumption that this torque was responsible for the Lehmann effect became a paradigm in the literature. Recent results, however, show that the Leslie coupling cannot be entirely responsible for the Lehmann effect, in particular because the first one is tied to the macroscopic chirality of the phase, independently of the macroscopic twist $q$ of the director field \cite{eber,dequidt1,oswald3,oswald4},  whereas the second one crucially depends on $q$. This was evidenced by the fact that the Leslie coupling is still observable at the compensation temperature of a cholesteric phase where $q=0$ \cite{eber,dequidt1,oswald3}, whereas the Lehmann rotation disappears at this point \cite{oswald5,oswald6}.  In addition, the Leslie torque and the rotation vector of the droplets are sometimes of opposite sign \cite{oswald4}.  As a consequence, these two effects must be clearly distinguished even if both seem to come from the chirality of the phase and the molecules.
}

An important question thus arises: can we observe the Lehmann and(or) Leslie effects in a phase --chiral or achiral-- of a material made of
achiral molecules? A part of the answer has been recently suggested by Brand {\it et al} \cite{brand}, who show theoretically that one might expect
thermomechanical effects similar to the Leslie coupling of CLC in chiral phases of achiral molecules, such as those formed by banana-shaped
molecules \cite{pelzl}. But this does not mean that a Lehmann effect can exist in these phases {\rev if the Leslie paradigm  is wrong}.
	
Here, we show that it is possible to observe the Lehmann rotation in a usual nematic LC, where both the molecules and the phase --of the ${\it
D}_{\infty h}$ symmetry-- are achiral. In the absence of molecular-level chiral induction, the required twist in the director field must be
provided by topological constraints. For example, this can be the case inside tangentially anchored bipolar nematic droplets provided that the
twist constant, $K_2$, is much smaller than the splay and bend constants, $K_1$ and $K_3$ \cite{williams}. Experimentally, this condition is
fulfilled in the nematic phase of the lyotropic LC made with aqueous suspensions of the food dye Sunset Yellow FCF (SSY) \cite{zhou},  where
twisted bipolar droplets nucleate in the coexistence region with the isotropic  liquid.  We recall that SSY is the commercial name of the disodium
salt of 6-hydroxy-5-[(4-sulfophenyl)azo]-2-naphtalenesulfonic acid. The compound was purchased from Sigma-Aldrich, and purified by successive
recrystallizations of saturated water solutions using ice-cold absolute ethanol. The wet precipitate was dried at 120$^\circ$C under vacuum during
a day. Aqueous solutions of SSY (molar mass 452.37) were prepared with concentrations typically ranging between 0.88 and 1.0 mol/kg, which results
in the nematic/isotropic coexistance above room temperature. Because of their flat central core of polyaromatic rings, SSY molecules tend to form
aggregates by face-to-face stacking in water. These elongated aggregates may self-assemble into ordered mesophases, constituting the so-called
lyotropic chromonic liquid crystals (LCLC) \cite{lavrentovich_bookLCLC}.

In our experiments, the LCLC was contained in a rectangular cell of gap 110 $\mu$m, whose glass plates were spin-coated with polyvinyl alcohol
(PVA) and annealed at 120$^\circ$C for 1 hour in order to favor wetting of the walls by the isotropic phase. The cell was placed between the two
transparent ovens of the experimental setup described in Ref.~\cite{oswald1}. A red bandpass filter was placed in front of the microscope light
source to optimize optical contrast and to minimize local heating of the sample. Two thin layers of glycerol ensured a good thermal contact between the cell and the ovens. The temperature gradient, perpendicular to the cell walls, is given by $G=\frac{\Delta T}{4e}\frac{\kappa_g}{\kappa_{LC}}$,
where $e=1~$mm is the thickness of the glass plates, $\Delta T=T_u-T_b$ is the temperature difference between the upper and bottom ovens and
$\kappa_g$ and $\kappa_{LC}$ are the thermal conductivities of the glass and the LC, respectively. By taking for $\kappa_{LC}$ the conductivity of
water, we obtain $\frac{\kappa_g}{\kappa_{LC}}\sim 1/0.6$, which gives $G$(K/m) $\sim 420\,\Delta T$~(K).
	
	\begin{figure}[t]
\includegraphics[width=8.5cm,clip]{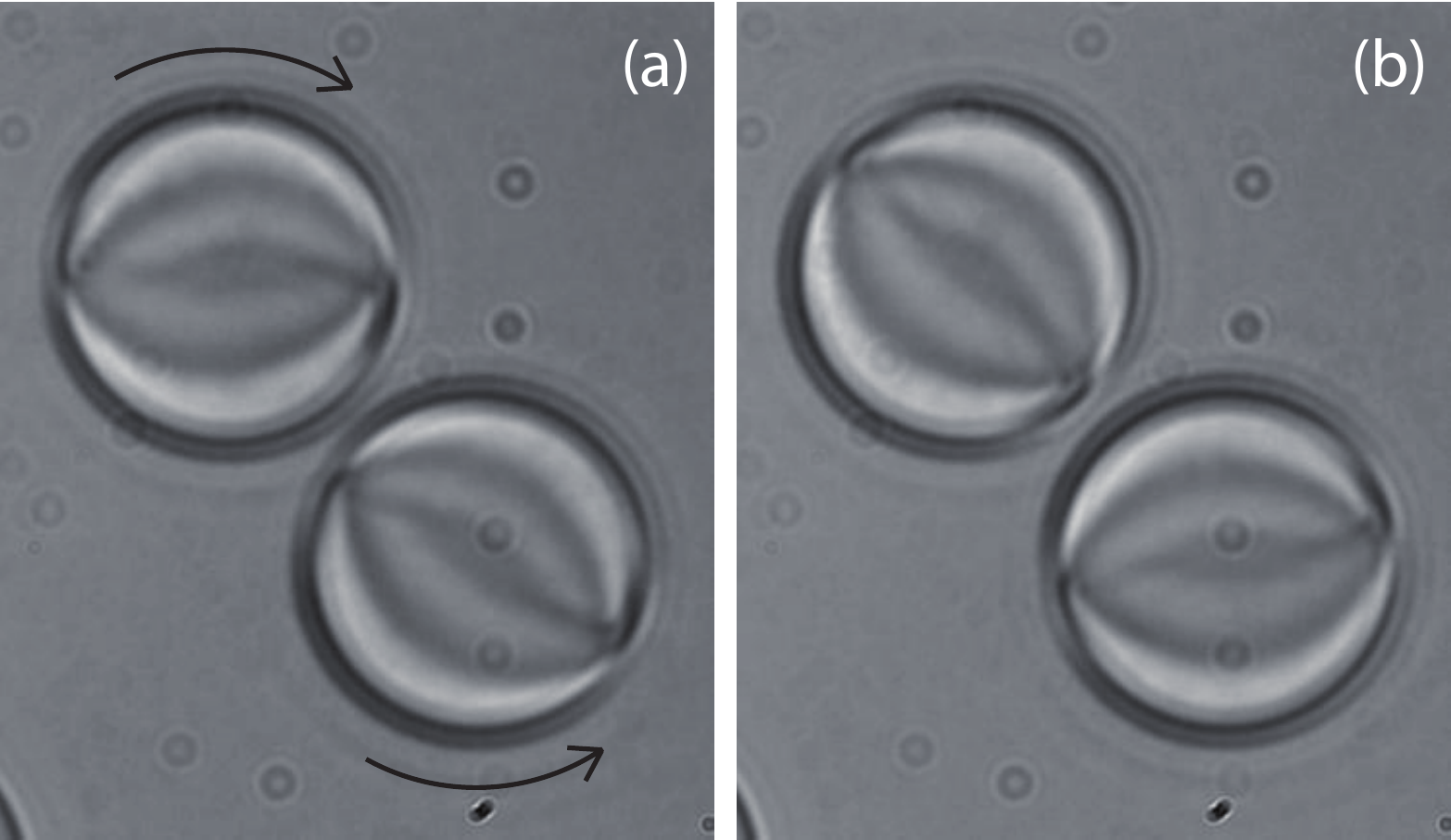}
\caption{\label{fig:rotating}  Optical micrograph of two twisted bipolar droplets of radius 25~$\mu$m rotating in opposite directions. a) $t=0$;
b) $t= 10s$. $T_{NI}= 45^\circ$C and $\Delta T=20^\circ$C. Images are taken using unpolarized red light illumination. The axis of the
droplets lays on the image plane.}
\end{figure}

	In order to observe the Lehmann rotation, the cell is heated in the nematic/isotropic coexistence region. In the studied LCLC, the {\rev coexistence} range is relatively large, of the order of 10-12$^\circ$C. Most of our observations are made at a temperature that falls roughly in the middle of the {\rev coexistence} range. Nematic droplets are at rest when they nucleate under uniform temperature conditions, and start to rotate when a temperature gradient is imposed, revealing the Lehmann effect. Contrary to what is observed in CLC, {\it we observe coexistence of droplets rotating in opposite senses}. This is visible in Fig.~\ref{fig:rotating}, and in Video S1  (scale bar 20 $\mu$m) of the supplementary material, where two neighboring droplets rotate with opposite handedness.
We argue that the rotation is due to the spontaneous twist of the droplet director field, which can be positive or negative with the same
likelihood, since the nematic phase is achiral.

Combined with the results reported for CLC, our experiments prove that the Lehmann rotation is due to the chirality of the director field and not
to the chirality of the phase, thus confirming that the Leslie mechanism, which is only possible in a chiral phase, cannot explain alone the
Lehmann rotation observed in CLC.

 \begin{figure}[t]
\includegraphics[width=8.5cm,clip]{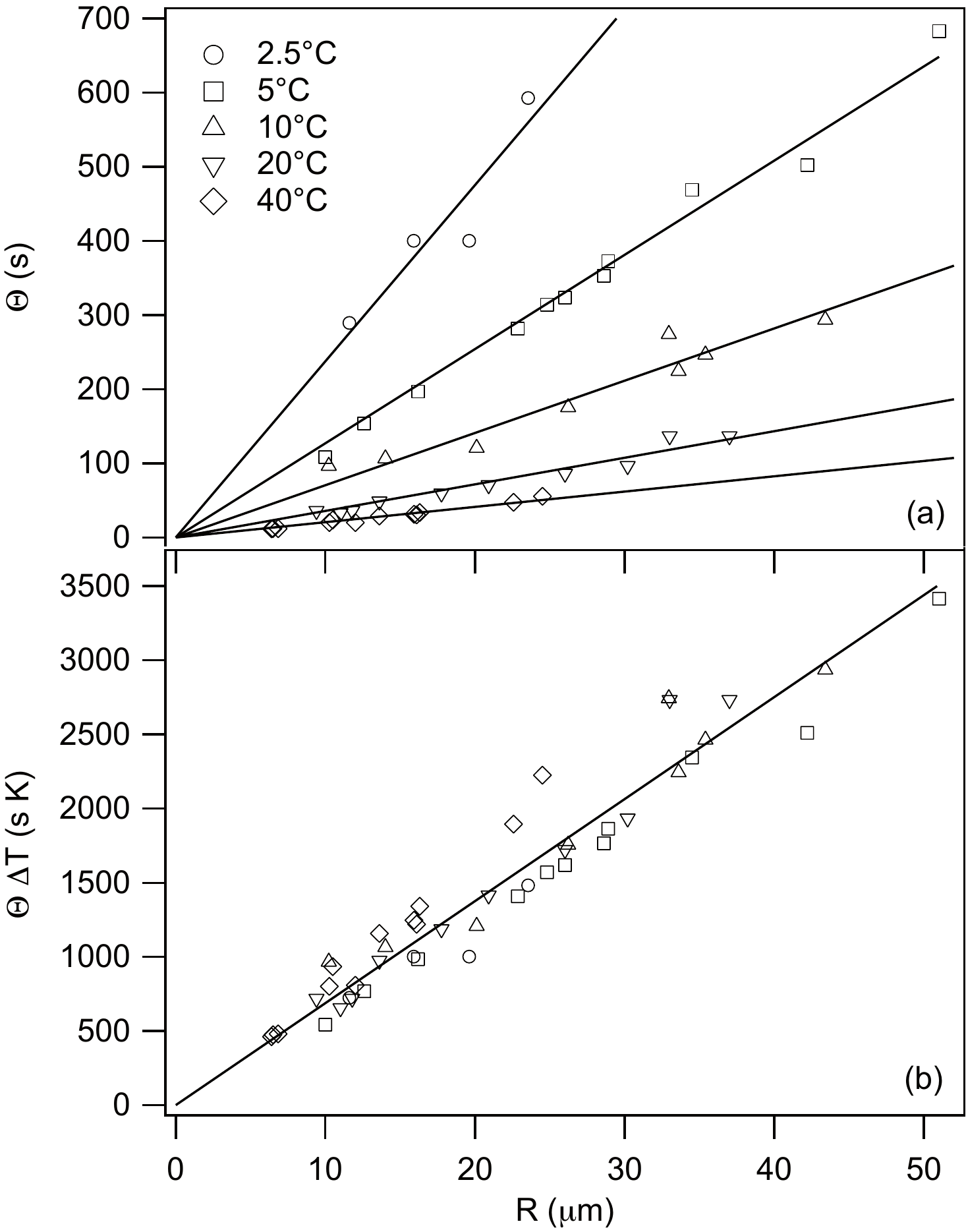}
\caption{\label{fig:data}  (a) Period of rotation, $\Theta$, as a function of droplet radius, $R$, for different temperature difference $\Delta
T$, and their fit to a straight line passing through the origin. (b) Product  $\Theta\Delta T$ as a function of $R$. $T_{NI} = 45^\circ$C. }
\end{figure}

\begin{figure}[t]
\includegraphics[width=8.5cm,clip]{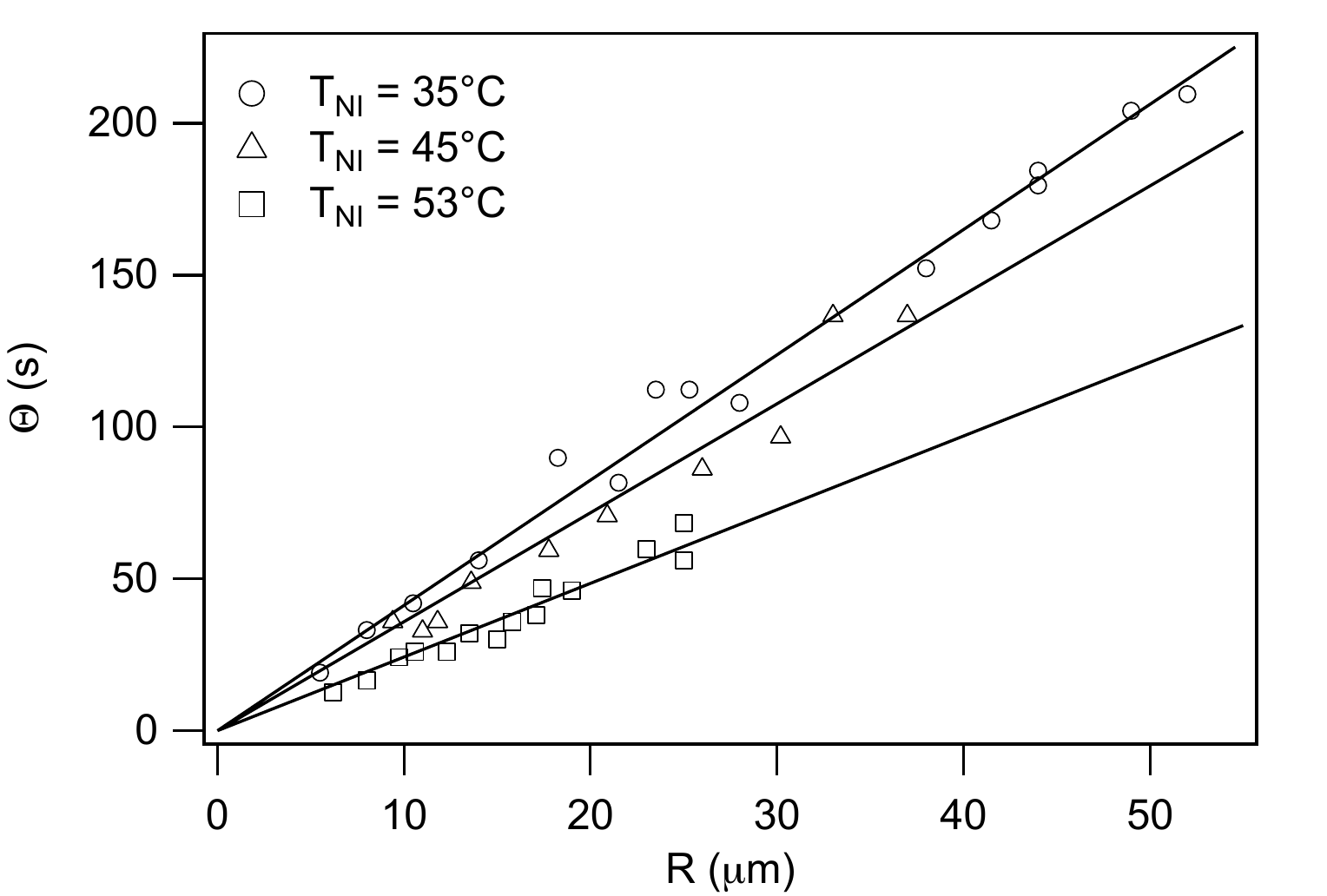}
\caption{\label{fig:moredata}  Period of rotation as a function of the radius at $\Delta T=20^\circ$C and different concentrations of SSY
corresponding to different $T_{NI}$. {\rev  {\large $\circ$}:   $T_{NI} = 35^\circ$C, SSY conc. 0.88 mol/kg; $\triangle$:  $T_{NI} = 45^\circ$C, SSY conc.  0.95 mol/kg; $\square$: $T_{NI} = 53^\circ$C, SSY conc. 1.0 mol/kg.} }
\end{figure}

To go further in the analysis of the phenomenon, we have explored the influence of the temperature gradient, the droplet radius, and the elastic
constants (through their temperature dependence) on the period of rotation, $\Theta$, of the nematic droplets.  Fig.~\ref{fig:data} shows data for
a sample with $T_{NI}\approx 45^\circ$C (SSY concentration $\approx 0.95$ mol/kg) in different temperature gradients. Here, $T_{NI}$ corresponds
to the center of the nematic/isotropic coexistence region. Temperature difference, $\Delta T$, ranges from $2.5^\circ$C until $40^\circ$C. For a
given $\Delta T$, the steady-state droplet size is increased (resp. decreased) by slightly decreasing (resp. increasing) the mean temperature
around $45^\circ$C, while maintaining a constant $\Delta T$.  Each dataset is fitted with a straight line with zero intercept. At least half a
turn is analyzed for each rotating droplet, and only droplets with a well-oriented bipolar configuration -- with the dipole axis perpendicular to
$\vec{G}$ --  are considered. We observe that higher $G$ favor droplets oriented with their axis either parallel or perpendicular to $\vec{G}$. At lower $G$, droplets with intermediate orientations are abundant but we do not analyze them because their tilt angle usually changes during the recording. Note that the droplets with their axis parallel to $\vec{G}$ do not rotate, which is expected since the director field is invariant under rotation about $\vec{G}$. We also note that large droplets are unstable with respect to coalescence in high $G$. This explains why the range of measured radii decreases when $G$ increases.  Fig.~\ref{fig:data}.a shows that $\Theta$ varies linearly with $R$ for each $\Delta T$, and
increases when $G$ decreases. Fig.~\ref{fig:data}.b shows that it is possible to collapse all these data on the same master
curve by plotting $\Theta\Delta T$ as a function of $R$. This proves that $\Theta$ is inversely proportional to $G$ or, equivalently, that the
rotation velocity is proportional to $G$. Finally, the period $\Theta$ for mixtures with different SSY concentrations, and so different
coexistence temperatures and elastic constants, is plotted in Fig.~\ref{fig:moredata}. Each dataset is taken at the same temperature gradient
($\Delta T=20^\circ$C).  This graph shows that the droplets slow down (the period of rotation increases) when the SSY concentration (or
equivalently, the transition temperature $T_{NI}$) decreases.	

In order to explain these data, we adapt the model of Ref.~\cite{oswald1}, originally developed for CLC, to the twisted bipolar nematic droplets.
In CLC, the driving torque for the rotation of the director is the Leslie thermomechanical torque given by {\rev $\vec{\Gamma}_{Leslie} = \nu \vec{n}\times (\vec{n}\times\vec{G})$, where $\vec{n}$ is the director, and $\nu$ is so-called the Leslie thermomechanical coefficient}. In the nematic case,
this torque vanishes for symmetry reasons ($\nu=0$). Nevertheless, Akopyan and Zeldovich have shown that, in deformed nematics, the director can still
experience a thermomechanical torque of general expression \cite{akopyan}:
  \begin{equation}\label{eq2}
\vec{\Gamma}_{TM} = \vec{n}\times \vec{f}_{TM},
\end{equation}
where the thermomechanical force is given by
\begin{eqnarray} \label{eq3}
\vec{f}_{TM}&=&\left(-\xi_1+\frac{\xi_3}{2}\right) (\vec{\nabla}\cdot\vec n) \;\vec G +\xi_2(\vec n \cdot \vec{\nabla}\times\vec n)( \vec n \times
\vec G)\nonumber \\
&+&\frac{1}{2}(\xi_3-\xi_4)(\vec n \times  (\vec{\nabla}\times\vec{n}))(\vec n \cdot \vec G)-\xi_3 \underline{\underline m} \vec G.
\end{eqnarray}
In this expression,  ${\underline{\underline m}}$ is the
symmetric tensor of components $m_{ij}=(n_{i,j}+n_{j,i})/2$, and  the  $\xi_i $'s (i=1...4) are phenomenological constants that account for the
different modes of director distortion that may give rise to thermomechanical torques. {\rev We emphasize that a more exact derivation of expression \eqref{eq3} has been given by Pleiner and Brand in Ref.\;\cite{pleiner}, p.\;44. In this work, the thermomechanical coefficients are denoted by $\Pi_i\;(i=1...4)$ with the correspondence $\xi_1 =  \gamma_1(\Pi_1+\Pi_2/2+\Pi_3/2), \xi_2 = \gamma_1(\Pi_2-\Pi_3)/2, \xi_3 = \gamma_1(\Pi_2+\Pi_3)$, and $\xi_4 = 2 \gamma_1\Pi_4$}. By assuming that there is no flow --- as in banded and
oriented droplets of CLC  \cite{poy}--- we obtain that the texture rotates at an angular velocity $\omega$ given by
  \begin{equation}\label{eq4}
\gamma_1\omega= \,\frac{\int_V \vec{f}_{TM}\cdot \vec{\delta}\,\dd ^3 \vec{r}}{\int_V \vec{\delta}^2\,\dd ^3 \vec{r}},
\end{equation}
where $V$ is the volume of the droplet and  $\vec \delta= \vec e_z \times \vec n - \frac{\partial \vec n}{\partial \vartheta}$. Here, $\vartheta$ is the polar angle in cylindrical coordinates centered on the droplet $(r, \vartheta, z)$, and the
unit vector $\vec e_z$ is parallel to the $z$ axis, which is perpendicular to the glass plates.

\begin{figure}[t]
\includegraphics[width=8.5cm,clip]{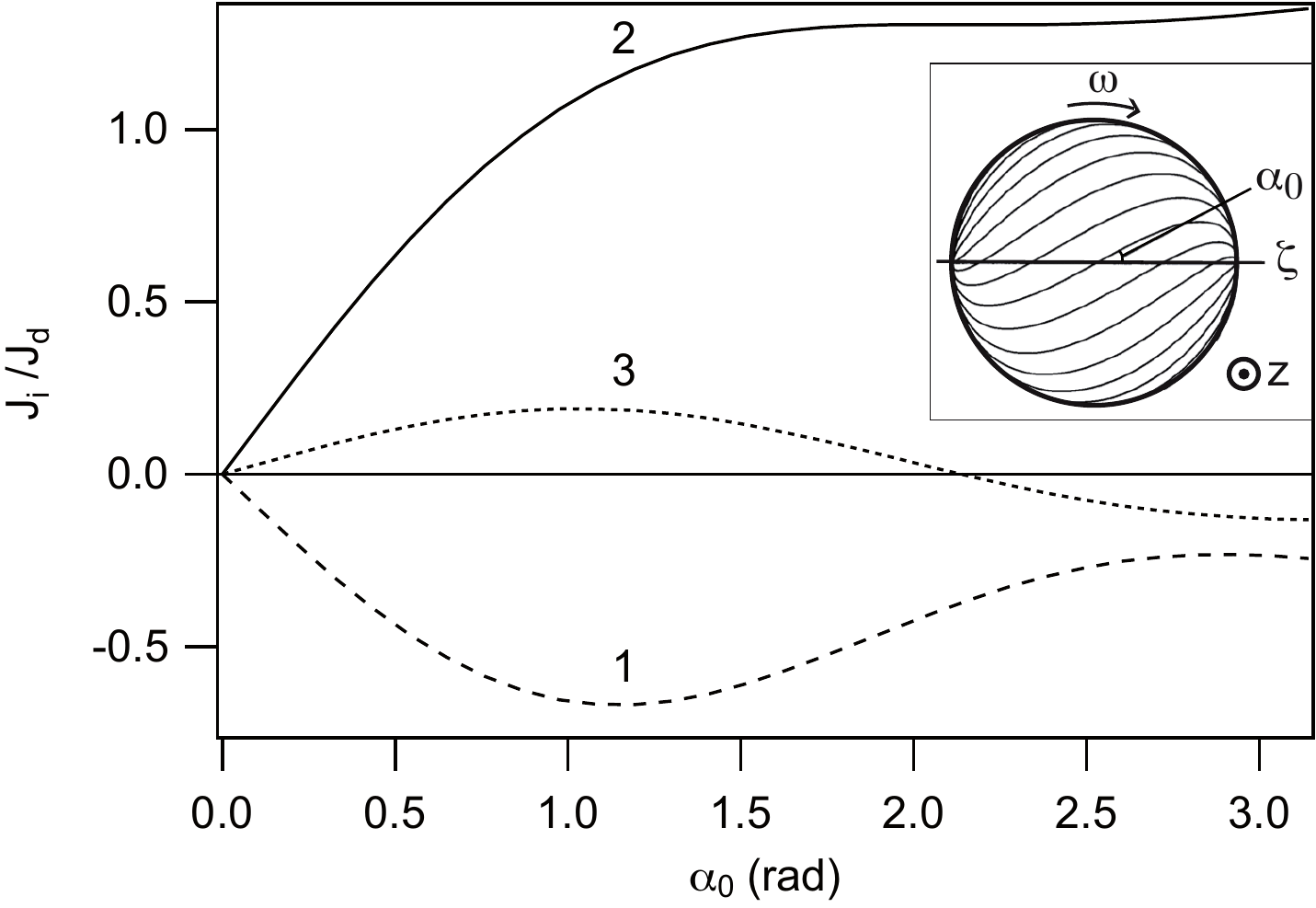}
\caption{\label{fig4}  Ratios of the splay $(i=1)$, twist $(i=2)$ and bend $(i=3)$ integrals $J_i$ over the dissipation integral $J_d$ as a function of the twist angle $\alpha_0$.  The value of $i$ is indicated besides each curve. Inset: angle $\alpha_0$ and director surface field lines in a twisted bipolar droplet. The $\zeta$ axis is the bipole axis and the $z$ axis is perpendicular to the glass plates and parallel to the temperature gradient.
}
\end{figure}

Expressions (\ref{eq3}) and  (\ref{eq4}) may be used to calculate the rotation velocity of the twisted bipolar droplets for a given configuration
of the director field inside the droplets. In order to simplify the calculations, we suppose that the planar anchoring at the droplet surface is
strong and use an ansatz for the director field that combines a pure bipolar configuration, $\vec{n}_b$, and a pure concentric configuration,
$\vec{n}_c$, namely $\vec{n}_{tb} = \vec{n}_b \cos (\alpha) - \vec{n}_c \sin(\alpha)$, with $\alpha(\rho)=\alpha_0 \rho/\rho_0$ \cite{xu,jeong}.
Here, $\rho$ is the radius in the cylindrical coordinates local to the droplet $(\rho, \varphi, \zeta)$, with the $\zeta$ axis taken along the
bipole , and $\rho_0=\sqrt{R^2-\zeta^2}$ ($(-R\le \zeta \le R)$. The angle $\alpha_0$ fixes the twist inside the droplet
(inset in Fig.~\ref{fig4}). This angle can be estimated by minimizing the Frank elastic energy of the droplet \cite{jeong}, and it is found
to only depend on the elastic anisotropy of the nematic phase, independently of the droplet radius. With typical values of
the elastic constants of the LCLC given in Ref.~\cite{zhou}, we
estimated  $\alpha_0 \sim 70^{\circ}$. {\rev We also measured $\alpha_0$ experimentally by analyzing the transmittance through the center of droplets, and considering a twisted planar nematic configuration in the Mauguin regime, in which the polarization of light follows the director \cite{jeong}. Using values for SSY birefringence reported in the literature \cite{horowitz2005}, our analysis yields $\alpha_0 = 100^{\circ}\pm 3^{\circ}$ in the range of concentrations used here, which is of the same order as the previous estimate}. Introducing the above approximation for the director field $\vec
n_{tb}$ in Eqs.~(\ref{eq2})-(\ref{eq4}) yields
\begin{equation}\label{eq5}
-\frac{\gamma_1 R \omega}{G}= \frac{\sum_i\bar \xi_i J_i(\alpha_0)}{J_d(\alpha_0)},
\end{equation}
with $\bar\xi_1=-\xi_1-\frac{\xi_3}{2}$,
$\bar\xi_2=\xi_2-\frac{\xi_3}{2}$, $\bar\xi_3=-\frac{\xi_4}{2}$, and

\begin{eqnarray} \label{eq6}
J_1&=&\int_{V_1}(\vec{\nabla}\cdot\vec n_{tb}) (\vec e_z\cdot\frac{\partial \vec n_{tb}}{\partial \vartheta})\,\dd^3 \vec{r}\,' \\
J_2&=&\int_{V_1} (\vec n_{tb} \cdot \vec{\nabla}\times \vec n_{tb}) [(\vec e_z\times \vec n_{tb})\cdot \vec{\delta}]\,\dd^3 \vec{r}\,'\\
J_3&=&\int_{V_1} (\vec n_{tb}\cdot\vec{e}_z)[\vec{n}\times(\vec{\nabla}\times \vec n_{tb})\cdot\vec{\delta}] \,\dd ^3 \vec{r}\,'\\
J_d&=&\int_{V_1}\vec{\delta}^2\,\dd^3 \vec{r}\,'.
\end{eqnarray}
Note that the  integrals $J_i$ ($J_d$), which are odd (even) functions of $\alpha_0$, are made dimensionless by using the variables $x'_i\equiv
x_i/R$ and are calculated over a spherical volume  $V_1$ of radius unity. The calculated ratios $J_i/J_d$, evaluated using Mathematica, are shown
in Fig.~\ref{fig4} as a function of the surface twist angle $\alpha_0$. They all vanish when $\alpha_0=0$, which means that $\omega=0$ when the
droplet is not twisted. This is expected  because, in that case, the droplet is superposable to its image in a mirror containing the temperature
gradient. But, since the droplet and its mirror image must have opposite velocities, this results in $\omega=-\omega=0$.  Eq. (\ref{eq5}) shows
that, in agreement with experiments, the rotation velocity $\omega$ is proportional to $G$ and inversely proportional to $R$. It must be emphasized that the dependence in $R$ arises from purely dimensional reasons in this model. Indeed, the only dimensionless quantity that can be formed with the
variables $R$, $G$, $\omega$, and the material constants $\gamma_1$ and $\bar\xi$ (with dimensions  N~m$^{-2}$~s and N~K$^{-1}$, respectively)  is
$\gamma_1 R\omega/(\bar\xi G)$. This result holds if $R$ is the only length scale in the problem, which supposes that the anchoring extrapolation length is much smaller than $R$ (strong anchoring). From our experimental data shown in Fig.~\ref{fig:data}.b, we can estimate a value for the ratio
$\bar\xi/\gamma_1$, by assuming that all the $\bar\xi_i$ are equal, and by taking $\alpha_0\sim 100^\circ$ we get
$\frac{\bar\xi}{\gamma_1} \approx 2.5~10^{-10}~$m$^2$~s$^{-1}$~K$^{-1}$. We measured the  rotational viscosity with a rotating magnetic field
according to the method described in Ref. \cite{oswald7} and by taking the value of the magnetic anisotropy given in Ref.~\cite{zhou} we get
$\gamma_1\approx0.32~$ Pa~s.  This gives $\bar\xi\approx
8~10^{-11}$~N~K$^{-1}$. We obtained the sign of $\bar\xi$ by doping the LCLC with a small
amount of {\rev the chiral dopant brucine sulphate. We verified,
by measuring the twist angle in cells with homogeneous planar
unidirectional (rubbed glass)/planar sliding (PVA coating) anchoring,
that this substance promotes a right-handed twist in SSY. In the
presence of this dopant}, most of the droplets rotate in the same
direction, with $\omega$ and $G$ of opposite signs, which leads to
$\bar\xi>0$.  We emphasize that our value of $\bar\xi$ is 50 times larger in absolute value than that measured experimentally in  hybrid layers of nematic LC by Lavrentovich and Nastishin \cite{lavrentovich} and  by Akopyan {\it et al.}
\cite{akopyan2}. This leads us to think that this model is, at most, qualitatively correct, and that another mechanism is  involved as in the
classical Lehmann effect.
Only independent and more precise measurements of the coefficients $\bar\xi_i$, for instance in the geometry proposed by Poursamad in
Ref.~\cite{poursamad}, will enable us to settle this issue.

In conclusion, the Lehmann effect also exists in usual nematic LC. In the experiments reported here, the rotation is due to a macroscopic chiral
symmetry breaking of the director field inside droplets with twisted bipolar director field. This configuration arises in SSY LCLC thanks to their
giant elastic anisotropy. The main difference with respect to CLC is that here, droplets rotating in both directions coexist in the same sample,
since their spontaneous twist develops indifferently to the right or to the left. This observation shows that the phase and the molecules do not
need to be chiral to observe a Lehmann rotation. This is different from the Leslie thermomechanical effect, which can only be observed in chiral
mesophases.  In the future, it would be interesting to measure more precisely the thermomechanical coefficients $\xi_i$ introduced by Akopyan and
Zel'dovich in usual nematics. Another question concerns the existence of similar terms in CLC. In principle, they should also exist in CLC but it
is not clear how to generalize  Eq.~\ref{eq3} to this phase. In particular, it is not clear {\rev whether} the thermomechanical force $\vec{f}_{TM}$ must be "regularized"  in order that it vanishes in a non-deformed cholesteric phase, the term in $\xi_3$ {\rev becoming} problematic. Once this theoretical problem is resolved, it would be interesting to check the role of these terms in the classical Lehmann effect. Nevertheless, {\rev we think} that the thermomechanical terms are not necessary to explain the Lehmann effect both in nematic and cholesteric LCs.

We thank A. Dequidt and E. Kats for fruitful discussions {\rev and B. Mart\'{\i}nez-Prat for assistance in the characterization of SSY samples.  J. I-M. acknowledges support by the CNRS for hosting his visit to the Physics laboratory of the ENS of Lyon. This work was partially suported by project MINECO-FIS2013-41144P (J. I-M.)}.

\bibliography{Ignes_Poy_Oswald_PRL2016cor3.4}

\begin{thebibliography}{41}%
\makeatletter
\providecommand \@ifxundefined [1]{%
 \@ifx{#1\undefined}
}%
\providecommand \@ifnum [1]{%
 \ifnum #1\expandafter \@firstoftwo
 \else \expandafter \@secondoftwo
 \fi
}%
\providecommand \@ifx [1]{%
 \ifx #1\expandafter \@firstoftwo
 \else \expandafter \@secondoftwo
 \fi
}%
\providecommand \natexlab [1]{#1}%
\providecommand \enquote  [1]{``#1''}%
\providecommand \bibnamefont  [1]{#1}%
\providecommand \bibfnamefont [1]{#1}%
\providecommand \citenamefont [1]{#1}%
\providecommand \href@noop [0]{\@secondoftwo}%
\providecommand \href [0]{\begingroup \@sanitize@url \@href}%
\providecommand \@href[1]{\@@startlink{#1}\@@href}%
\providecommand \@@href[1]{\endgroup#1\@@endlink}%
\providecommand \@sanitize@url [0]{\catcode `\\12\catcode `\$12\catcode
  `\&12\catcode `\#12\catcode `\^12\catcode `\_12\catcode `\%12\relax}%
\providecommand \@@startlink[1]{}%
\providecommand \@@endlink[0]{}%
\providecommand \url  [0]{\begingroup\@sanitize@url \@url }%
\providecommand \@url [1]{\endgroup\@href {#1}{\urlprefix }}%
\providecommand \urlprefix  [0]{URL }%
\providecommand \Eprint [0]{\href }%
\providecommand \doibase [0]{http://dx.doi.org/}%
\providecommand \selectlanguage [0]{\@gobble}%
\providecommand \bibinfo  [0]{\@secondoftwo}%
\providecommand \bibfield  [0]{\@secondoftwo}%
\providecommand \translation [1]{[#1]}%
\providecommand \BibitemOpen [0]{}%
\providecommand \bibitemStop [0]{}%
\providecommand \bibitemNoStop [0]{.\EOS\space}%
\providecommand \EOS [0]{\spacefactor3000\relax}%
\providecommand \BibitemShut  [1]{\csname bibitem#1\endcsname}%
\let\auto@bib@innerbib\@empty
\bibitem [{\citenamefont {Milch}\ and\ \citenamefont {Sykes}(2009)}]{michl}%
  \BibitemOpen
  \bibfield  {author} {\bibinfo {author} {\bibfnamefont {J.}~\bibnamefont
  {Milch}}\ and\ \bibinfo {author} {\bibfnamefont {E.~C.~H.}\ \bibnamefont
  {Sykes}},\ }\href@noop {} {\bibfield  {journal} {\bibinfo  {journal} {ACS
  Nano}\ }\textbf {\bibinfo {volume} {3}},\ \bibinfo {pages} {1042} (\bibinfo
  {year} {2009})}\BibitemShut {NoStop}%
\bibitem [{\citenamefont {de~Gennes}(1974)}]{degennes}%
  \BibitemOpen
  \bibfield  {author} {\bibinfo {author} {\bibfnamefont {P.-G.}\ \bibnamefont
  {de~Gennes}},\ }\href@noop {} {\emph {\bibinfo {title} {The {Physics} of
  {Liquid} {Crystals}}}}\ (\bibinfo  {publisher} {Oxford University Press,
  USA},\ \bibinfo {year} {1974})\BibitemShut {NoStop}%
\bibitem [{\citenamefont {Tabe}\ and\ \citenamefont {Yokoyama}(2003)}]{tabe2}%
  \BibitemOpen
  \bibfield  {author} {\bibinfo {author} {\bibfnamefont {Y.}~\bibnamefont
  {Tabe}}\ and\ \bibinfo {author} {\bibfnamefont {H.}~\bibnamefont
  {Yokoyama}},\ }\href@noop {} {\bibfield  {journal} {\bibinfo  {journal}
  {Nature Mat.}\ }\textbf {\bibinfo {volume} {2}},\ \bibinfo {pages} {806}
  (\bibinfo {year} {2003})}\BibitemShut {NoStop}%
\bibitem [{\citenamefont {Milczarczyk-Piwowarczyk}\ \emph
  {et~al.}(2008)\citenamefont {Milczarczyk-Piwowarczyk}, \citenamefont
  {\.Zywoci\'nski}, \citenamefont {Noworyta},\ and\ \citenamefont
  {Holyst}}]{zywo}%
  \BibitemOpen
  \bibfield  {author} {\bibinfo {author} {\bibfnamefont {P.}~\bibnamefont
  {Milczarczyk-Piwowarczyk}}, \bibinfo {author} {\bibfnamefont
  {A.}~\bibnamefont {\.Zywoci\'nski}}, \bibinfo {author} {\bibfnamefont
  {K.}~\bibnamefont {Noworyta}}, \ and\ \bibinfo {author} {\bibfnamefont
  {R.}~\bibnamefont {Holyst}},\ }\href@noop {} {\bibfield  {journal} {\bibinfo
  {journal} {Langmuir}\ }\textbf {\bibinfo {volume} {24}},\ \bibinfo {pages}
  {12354} (\bibinfo {year} {2008})}\BibitemShut {NoStop}%
\bibitem [{\citenamefont {Seki}\ \emph {et~al.}(2011)\citenamefont {Seki},
  \citenamefont {Ueda}, \citenamefont {Okumura},\ and\ \citenamefont
  {Tabe}}]{seki}%
  \BibitemOpen
  \bibfield  {author} {\bibinfo {author} {\bibfnamefont {K.}~\bibnamefont
  {Seki}}, \bibinfo {author} {\bibfnamefont {K.}~\bibnamefont {Ueda}}, \bibinfo
  {author} {\bibfnamefont {Y.-i.}\ \bibnamefont {Okumura}}, \ and\ \bibinfo
  {author} {\bibfnamefont {Y.}~\bibnamefont {Tabe}},\ }\href@noop {} {\bibfield
   {journal} {\bibinfo  {journal} {J. Phys. Condens. Matter}\ }\textbf
  {\bibinfo {volume} {23}},\ \bibinfo {pages} {284114} (\bibinfo {year}
  {2011})}\BibitemShut {NoStop}%
\bibitem [{\citenamefont {Gil}\ and\ \citenamefont {Thiberge}(1997)}]{gil1}%
  \BibitemOpen
  \bibfield  {author} {\bibinfo {author} {\bibfnamefont {L.}~\bibnamefont
  {Gil}}\ and\ \bibinfo {author} {\bibfnamefont {S.}~\bibnamefont {Thiberge}},\
  }\href@noop {} {\bibfield  {journal} {\bibinfo  {journal} {J. Phys. II
  France}\ }\textbf {\bibinfo {volume} {7}},\ \bibinfo {pages} {1499} (\bibinfo
  {year} {1997})}\BibitemShut {NoStop}%
\bibitem [{\citenamefont {Madhusudana}\ and\ \citenamefont
  {Pratibha}(1987)}]{madhusudana}%
  \BibitemOpen
  \bibfield  {author} {\bibinfo {author} {\bibfnamefont {N.~V.}\ \bibnamefont
  {Madhusudana}}\ and\ \bibinfo {author} {\bibfnamefont {R.}~\bibnamefont
  {Pratibha}},\ }\href@noop {} {\bibfield  {journal} {\bibinfo  {journal} {Mol.
  Cryst. Liq. Cryst.}\ }\textbf {\bibinfo {volume} {5}},\ \bibinfo {pages} {43}
  (\bibinfo {year} {1987})}\BibitemShut {NoStop}%
\bibitem [{\citenamefont {Padmini}\ and\ \citenamefont
  {Madhusudana}(1993)}]{padmini}%
  \BibitemOpen
  \bibfield  {author} {\bibinfo {author} {\bibfnamefont {H.~P.}\ \bibnamefont
  {Padmini}}\ and\ \bibinfo {author} {\bibfnamefont {N.~V.}\ \bibnamefont
  {Madhusudana}},\ }\href@noop {} {\bibfield  {journal} {\bibinfo  {journal}
  {Liq. Cryst.}\ }\textbf {\bibinfo {volume} {5}},\ \bibinfo {pages} {497}
  (\bibinfo {year} {1993})}\BibitemShut {NoStop}%
\bibitem [{\citenamefont {Gil}\ and\ \citenamefont {Gilli}(1998)}]{gil2}%
  \BibitemOpen
  \bibfield  {author} {\bibinfo {author} {\bibfnamefont {L.}~\bibnamefont
  {Gil}}\ and\ \bibinfo {author} {\bibfnamefont {J.~M.}\ \bibnamefont
  {Gilli}},\ }\href@noop {} {\bibfield  {journal} {\bibinfo  {journal} {Phys.
  Rev. Lett.}\ }\textbf {\bibinfo {volume} {80}},\ \bibinfo {pages} {5742}
  (\bibinfo {year} {1998})}\BibitemShut {NoStop}%
\bibitem [{\citenamefont {Baudry}\ \emph {et~al.}(1999)\citenamefont {Baudry},
  \citenamefont {Pirkl},\ and\ \citenamefont {Oswald}}]{baudry}%
  \BibitemOpen
  \bibfield  {author} {\bibinfo {author} {\bibfnamefont {J.}~\bibnamefont
  {Baudry}}, \bibinfo {author} {\bibfnamefont {S.}~\bibnamefont {Pirkl}}, \
  and\ \bibinfo {author} {\bibfnamefont {P.}~\bibnamefont {Oswald}},\
  }\href@noop {} {\bibfield  {journal} {\bibinfo  {journal} {Phys. Rev. E}\
  }\textbf {\bibinfo {volume} {57}},\ \bibinfo {pages} {2990} (\bibinfo {year}
  {1999})}\BibitemShut {NoStop}%
\bibitem [{\citenamefont {Tarasov}\ \emph {et~al.}(2003)\citenamefont
  {Tarasov}, \citenamefont {Krekhov},\ and\ \citenamefont {Kramer}}]{kramer}%
  \BibitemOpen
  \bibfield  {author} {\bibinfo {author} {\bibfnamefont {O.~S.}\ \bibnamefont
  {Tarasov}}, \bibinfo {author} {\bibfnamefont {A.~P.}\ \bibnamefont
  {Krekhov}}, \ and\ \bibinfo {author} {\bibfnamefont {L.}~\bibnamefont
  {Kramer}},\ }\href@noop {} {\bibfield  {journal} {\bibinfo  {journal} {Phys.
  Rev. E}\ }\textbf {\bibinfo {volume} {68}},\ \bibinfo {pages} {031708}
  (\bibinfo {year} {2003})}\BibitemShut {NoStop}%
\bibitem [{\citenamefont {Tarasov}(2003)}]{tarasov}%
  \BibitemOpen
  \bibfield  {author} {\bibinfo {author} {\bibfnamefont {O.~S.}\ \bibnamefont
  {Tarasov}},\ }\emph {\bibinfo {title} {Structural Transition and Dynamics of
  Liquid Crystals under Flows and Electric Fields}},\ \href@noop {} {Ph.D.
  thesis},\ \bibinfo  {school} {University of Bayreuth} (\bibinfo {year}
  {2003})\BibitemShut {NoStop}%
\bibitem [{\citenamefont {Dequidt}\ and\ \citenamefont
  {Oswald}(2007{\natexlab{a}})}]{dequidt}%
  \BibitemOpen
  \bibfield  {author} {\bibinfo {author} {\bibfnamefont {A.}~\bibnamefont
  {Dequidt}}\ and\ \bibinfo {author} {\bibfnamefont {P.}~\bibnamefont
  {Oswald}},\ }\href@noop {} {\bibfield  {journal} {\bibinfo  {journal} {Eur.
  Phys. J. E}\ }\textbf {\bibinfo {volume} {24}},\ \bibinfo {pages} {157}
  (\bibinfo {year} {2007}{\natexlab{a}})}\BibitemShut {NoStop}%
\bibitem [{\citenamefont {Lehmann}(1900)}]{lehmann}%
  \BibitemOpen
  \bibfield  {author} {\bibinfo {author} {\bibfnamefont {O.}~\bibnamefont
  {Lehmann}},\ }\href@noop {} {\bibfield  {journal} {\bibinfo  {journal} {Ann.\
  Phys.}\ }\textbf {\bibinfo {volume} {2}},\ \bibinfo {pages} {649} (\bibinfo
  {year} {1900})}\BibitemShut {NoStop}%
\bibitem [{\citenamefont {Oswald}\ and\ \citenamefont
  {Dequidt}(2008{\natexlab{a}})}]{oswald1}%
  \BibitemOpen
  \bibfield  {author} {\bibinfo {author} {\bibfnamefont {P.}~\bibnamefont
  {Oswald}}\ and\ \bibinfo {author} {\bibfnamefont {A.}~\bibnamefont
  {Dequidt}},\ }\href@noop {} {\bibfield  {journal} {\bibinfo  {journal} {Phys.
  Rev. Lett.}\ }\textbf {\bibinfo {volume} {100}},\ \bibinfo {pages} {217802}
  (\bibinfo {year} {2008}{\natexlab{a}})}\BibitemShut {NoStop}%
\bibitem [{\citenamefont {Oswald}(2009)}]{oswald2}%
  \BibitemOpen
  \bibfield  {author} {\bibinfo {author} {\bibfnamefont {P.}~\bibnamefont
  {Oswald}},\ }\href@noop {} {\bibfield  {journal} {\bibinfo  {journal} {Eur.
  Phys. J. E}\ }\textbf {\bibinfo {volume} {28}},\ \bibinfo {pages} {377}
  (\bibinfo {year} {2009})}\BibitemShut {NoStop}%
\bibitem [{\citenamefont {Yoshioka}\ \emph {et~al.}(2014)\citenamefont
  {Yoshioka}, \citenamefont {Ito}, \citenamefont {Suzuki}, \citenamefont
  {Takahashi}, \citenamefont {Takizawa},\ and\ \citenamefont {Tabe}}]{tabe}%
  \BibitemOpen
  \bibfield  {author} {\bibinfo {author} {\bibfnamefont {J.}~\bibnamefont
  {Yoshioka}}, \bibinfo {author} {\bibfnamefont {F.}~\bibnamefont {Ito}},
  \bibinfo {author} {\bibfnamefont {Y.}~\bibnamefont {Suzuki}}, \bibinfo
  {author} {\bibfnamefont {H.}~\bibnamefont {Takahashi}}, \bibinfo {author}
  {\bibfnamefont {H.}~\bibnamefont {Takizawa}}, \ and\ \bibinfo {author}
  {\bibfnamefont {Y.}~\bibnamefont {Tabe}},\ }\href@noop {} {\bibfield
  {journal} {\bibinfo  {journal} {Soft Matter}\ }\textbf {\bibinfo {volume}
  {10}},\ \bibinfo {pages} {5869} (\bibinfo {year} {2014})}\BibitemShut
  {NoStop}%
\bibitem [{\citenamefont {Yamamoto}\ \emph {et~al.}(2006)\citenamefont
  {Yamamoto}, \citenamefont {Kuroda},\ and\ \citenamefont {Sano}}]{sano}%
  \BibitemOpen
  \bibfield  {author} {\bibinfo {author} {\bibfnamefont {T.}~\bibnamefont
  {Yamamoto}}, \bibinfo {author} {\bibfnamefont {M.}~\bibnamefont {Kuroda}}, \
  and\ \bibinfo {author} {\bibfnamefont {M.}~\bibnamefont {Sano}},\ }\href@noop
  {} {\bibfield  {journal} {\bibinfo  {journal} {Europhys. Lett.}\ }\textbf
  {\bibinfo {volume} {109}},\ \bibinfo {pages} {46001} (\bibinfo {year}
  {2006})}\BibitemShut {NoStop}%
\bibitem [{\citenamefont {Leslie}(1968)}]{leslie1}%
  \BibitemOpen
  \bibfield  {author} {\bibinfo {author} {\bibfnamefont {F.~M.}\ \bibnamefont
  {Leslie}},\ }\href@noop {} {\bibfield  {journal} {\bibinfo  {journal} {Arch.
  Rational Mech. Anal.}\ }\textbf {\bibinfo {volume} {28}},\ \bibinfo {pages}
  {265} (\bibinfo {year} {1968})}\BibitemShut {NoStop}%
\bibitem [{\citenamefont {Leslie}(1971)}]{leslie2}%
  \BibitemOpen
  \bibfield  {author} {\bibinfo {author} {\bibfnamefont {F.~M.}\ \bibnamefont
  {Leslie}},\ }\href@noop {} {\bibfield  {journal} {\bibinfo  {journal} {Symp.
  Faraday Soc.}\ }\textbf {\bibinfo {volume} {5}},\ \bibinfo {pages} {33}
  (\bibinfo {year} {1971})}\BibitemShut {NoStop}%
\bibitem [{\citenamefont {\'Eber}\ and\ \citenamefont
  {J\'anossy}(1982)}]{eber}%
  \BibitemOpen
  \bibfield  {author} {\bibinfo {author} {\bibfnamefont {N.}~\bibnamefont
  {\'Eber}}\ and\ \bibinfo {author} {\bibfnamefont {I.}~\bibnamefont
  {J\'anossy}},\ }\href@noop {} {\bibfield  {journal} {\bibinfo  {journal}
  {Mol.\ Cryst.\ Liq.\ Cryst.\ Lett.}\ }\textbf {\bibinfo {volume} {72}},\
  \bibinfo {pages} {233} (\bibinfo {year} {1982})}\BibitemShut {NoStop}%
\bibitem [{\citenamefont {Dequidt}\ and\ \citenamefont
  {Oswald}(2007{\natexlab{b}})}]{dequidt1}%
  \BibitemOpen
  \bibfield  {author} {\bibinfo {author} {\bibfnamefont {A.}~\bibnamefont
  {Dequidt}}\ and\ \bibinfo {author} {\bibfnamefont {P.}~\bibnamefont
  {Oswald}},\ }\href@noop {} {\bibfield  {journal} {\bibinfo  {journal}
  {Europhys. Lett.}\ }\textbf {\bibinfo {volume} {80}},\ \bibinfo {pages}
  {26001} (\bibinfo {year} {2007}{\natexlab{b}})}\BibitemShut {NoStop}%
\bibitem [{\citenamefont {Oswald}\ and\ \citenamefont
  {Dequidt}(2008{\natexlab{b}})}]{oswald3}%
  \BibitemOpen
  \bibfield  {author} {\bibinfo {author} {\bibfnamefont {P.}~\bibnamefont
  {Oswald}}\ and\ \bibinfo {author} {\bibfnamefont {A.}~\bibnamefont
  {Dequidt}},\ }\href@noop {} {\bibfield  {journal} {\bibinfo  {journal}
  {Europhys. Lett.}\ }\textbf {\bibinfo {volume} {83}},\ \bibinfo {pages}
  {16005} (\bibinfo {year} {2008}{\natexlab{b}})}\BibitemShut {NoStop}%
\bibitem [{\citenamefont {Oswald}(2014)}]{oswald4}%
  \BibitemOpen
  \bibfield  {author} {\bibinfo {author} {\bibfnamefont {P.}~\bibnamefont
  {Oswald}},\ }\href@noop {} {\bibfield  {journal} {\bibinfo  {journal}
  {Europhys. Lett.}\ }\textbf {\bibinfo {volume} {108}},\ \bibinfo {pages}
  {36001 and 59001 (Erratum)} (\bibinfo {year} {2014})}\BibitemShut {NoStop}%
\bibitem [{\citenamefont {Oswald}\ \emph {et~al.}(2011)\citenamefont {Oswald},
  \citenamefont {J\/orgensen},\ and\ \citenamefont {\.Zywoci\'nski}}]{oswald5}%
  \BibitemOpen
  \bibfield  {author} {\bibinfo {author} {\bibfnamefont {P.}~\bibnamefont
  {Oswald}}, \bibinfo {author} {\bibfnamefont {L.}~\bibnamefont {J\/orgensen}},
  \ and\ \bibinfo {author} {\bibfnamefont {A.}~\bibnamefont {\.Zywoci\'nski}},\
  }\href@noop {} {\bibfield  {journal} {\bibinfo  {journal} {Liq. Cryst.}\
  }\textbf {\bibinfo {volume} {38}},\ \bibinfo {pages} {601} (\bibinfo {year}
  {2011})}\BibitemShut {NoStop}%
\bibitem [{\citenamefont {Oswald}(2012{\natexlab{a}})}]{oswald6}%
  \BibitemOpen
  \bibfield  {author} {\bibinfo {author} {\bibfnamefont {P.}~\bibnamefont
  {Oswald}},\ }\href@noop {} {\bibfield  {journal} {\bibinfo  {journal} {Eur.
  Phys. J. E}\ }\textbf {\bibinfo {volume} {35}},\ \bibinfo {pages} {10}
  (\bibinfo {year} {2012}{\natexlab{a}})}\BibitemShut {NoStop}%
\bibitem [{\citenamefont {Brand}\ \emph {et~al.}(2012)\citenamefont {Brand},
  \citenamefont {Pleiner},\ and\ \citenamefont {Sven$\check{s}$ek}}]{brand}%
  \BibitemOpen
  \bibfield  {author} {\bibinfo {author} {\bibfnamefont {H.}~\bibnamefont
  {Brand}}, \bibinfo {author} {\bibfnamefont {H.}~\bibnamefont {Pleiner}}, \
  and\ \bibinfo {author} {\bibfnamefont {D.}~\bibnamefont
  {Sven$\check{s}$ek}},\ }\href@noop {} {\bibfield  {journal} {\bibinfo
  {journal} {Phys. Rev. E}\ }\textbf {\bibinfo {volume} {88}},\ \bibinfo
  {pages} {024501} (\bibinfo {year} {2012})}\BibitemShut {NoStop}%
\bibitem [{\citenamefont {Pelzl}\ \emph {et~al.}(1999)\citenamefont {Pelzl},
  \citenamefont {Diele},\ and\ \citenamefont {Weissflog}}]{pelzl}%
  \BibitemOpen
  \bibfield  {author} {\bibinfo {author} {\bibfnamefont {G.}~\bibnamefont
  {Pelzl}}, \bibinfo {author} {\bibfnamefont {S.}~\bibnamefont {Diele}}, \ and\
  \bibinfo {author} {\bibfnamefont {W.}~\bibnamefont {Weissflog}},\ }\href@noop
  {} {\bibfield  {journal} {\bibinfo  {journal} {Adv. Mat.}\ }\textbf {\bibinfo
  {volume} {11}},\ \bibinfo {pages} {707} (\bibinfo {year} {1999})}\BibitemShut
  {NoStop}%
\bibitem [{\citenamefont {Williams}(2012)}]{williams}%
  \BibitemOpen
  \bibfield  {author} {\bibinfo {author} {\bibfnamefont {R.~D.}\ \bibnamefont
  {Williams}},\ }\href@noop {} {\bibfield  {journal} {\bibinfo  {journal} {J.
  Phys. Math. Gen.}\ }\textbf {\bibinfo {volume} {19}},\ \bibinfo {pages}
  {3211} (\bibinfo {year} {2012})}\BibitemShut {NoStop}%
\bibitem [{\citenamefont {Zhou}\ and\ \citenamefont {co~workers}(2012)}]{zhou}%
  \BibitemOpen
  \bibfield  {author} {\bibinfo {author} {\bibfnamefont {S.}~\bibnamefont
  {Zhou}}\ and\ \bibinfo {author} {\bibnamefont {co~workers}},\ }\href@noop {}
  {\bibfield  {journal} {\bibinfo  {journal} {Phys. Rev. Lett.}\ }\textbf
  {\bibinfo {volume} {109}},\ \bibinfo {pages} {037801} (\bibinfo {year}
  {2012})}\BibitemShut {NoStop}%
\bibitem [{\citenamefont {Park}\ and\ \citenamefont
  {Lavrentovich}(2012)}]{lavrentovich_bookLCLC}%
  \BibitemOpen
  \bibfield  {author} {\bibinfo {author} {\bibfnamefont {H.-S.}\ \bibnamefont
  {Park}}\ and\ \bibinfo {author} {\bibfnamefont {O.}~\bibnamefont
  {Lavrentovich}},\ }\enquote {\bibinfo {title} {Lyotropic chromonic liquid
  crystals: Emerging applications},}\ in\ \href@noop {} {\emph {\bibinfo
  {booktitle} {Liquid Crystals Beyond Displays: Chemistry, Physics, and
  Applications,}}},\ \bibinfo {editor} {edited by\ \bibinfo {editor}
  {\bibfnamefont {Q.}~\bibnamefont {Li}}}\ (\bibinfo  {publisher} {John Wiley
  \& Sons, Inc.},\ \bibinfo {year} {2012})\ Chap.~\bibinfo {chapter}
  {14}\BibitemShut {NoStop}%
\bibitem [{\citenamefont {Akopyan}\ and\ \citenamefont
  {Zel'dovich}(1984)}]{akopyan}%
  \BibitemOpen
  \bibfield  {author} {\bibinfo {author} {\bibfnamefont {R.~S.}\ \bibnamefont
  {Akopyan}}\ and\ \bibinfo {author} {\bibfnamefont {B.~Y.}\ \bibnamefont
  {Zel'dovich}},\ }\href@noop {} {\bibfield  {journal} {\bibinfo  {journal}
  {Sov. Phys. JETP}\ }\textbf {\bibinfo {volume} {60}},\ \bibinfo {pages} {953}
  (\bibinfo {year} {1984})}\BibitemShut {NoStop}%
\bibitem [{\citenamefont {Pleiner}\ and\ \citenamefont
  {Brand}(1996)}]{pleiner}%
  \BibitemOpen
  \bibfield  {author} {\bibinfo {author} {\bibfnamefont {H.}~\bibnamefont
  {Pleiner}}\ and\ \bibinfo {author} {\bibfnamefont {H.~R.}\ \bibnamefont
  {Brand}},\ }in\ \href@noop {} {\emph {\bibinfo {booktitle} {Pattern Formation
  in Liquid Crystals}}}\ (\bibinfo  {publisher} {Springer},\ \bibinfo {year}
  {1996})\ pp.\ \bibinfo {pages} {15--67}\BibitemShut {NoStop}%
\bibitem [{\citenamefont {Poy}\ and\ \citenamefont {Oswald}(2016)}]{poy}%
  \BibitemOpen
  \bibfield  {author} {\bibinfo {author} {\bibfnamefont {G.}~\bibnamefont
  {Poy}}\ and\ \bibinfo {author} {\bibfnamefont {P.}~\bibnamefont {Oswald}},\
  }\href@noop {} {\bibfield  {journal} {\bibinfo  {journal} {Soft Matter}\
  }\textbf {\bibinfo {volume} {12}},\ \bibinfo {pages} {2604} (\bibinfo {year}
  {2016})}\BibitemShut {NoStop}%
\bibitem [{\citenamefont {Xu}\ and\ \citenamefont {Crooker}(1997)}]{xu}%
  \BibitemOpen
  \bibfield  {author} {\bibinfo {author} {\bibfnamefont {F.}~\bibnamefont
  {Xu}}\ and\ \bibinfo {author} {\bibfnamefont {P.}~\bibnamefont {Crooker}},\
  }\href@noop {} {\bibfield  {journal} {\bibinfo  {journal} {Phys. Rev. E}\
  }\textbf {\bibinfo {volume} {56}},\ \bibinfo {pages} {6853} (\bibinfo {year}
  {1997})}\BibitemShut {NoStop}%
\bibitem [{\citenamefont {Jeong}\ \emph {et~al.}(2014)\citenamefont {Jeong},
  \citenamefont {Davidson}, \citenamefont {Collings}, \citenamefont
  {Lubensky},\ and\ \citenamefont {Yodh}}]{jeong}%
  \BibitemOpen
  \bibfield  {author} {\bibinfo {author} {\bibfnamefont {J.}~\bibnamefont
  {Jeong}}, \bibinfo {author} {\bibfnamefont {Z.~S.}\ \bibnamefont {Davidson}},
  \bibinfo {author} {\bibfnamefont {P.~J.}\ \bibnamefont {Collings}}, \bibinfo
  {author} {\bibfnamefont {T.~C.}\ \bibnamefont {Lubensky}}, \ and\ \bibinfo
  {author} {\bibfnamefont {A.~G.}\ \bibnamefont {Yodh}},\ }\href@noop {}
  {\bibfield  {journal} {\bibinfo  {journal} {Proc. Natl. Acad. Sci. U S A}\
  }\textbf {\bibinfo {volume} {111}},\ \bibinfo {pages} {1742} (\bibinfo {year}
  {2014})}\BibitemShut {NoStop}%
\bibitem [{\citenamefont {Horowitz}\ \emph {et~al.}(2005)\citenamefont
  {Horowitz}, \citenamefont {Janowitz}, \citenamefont {Modic}, \citenamefont
  {Heiney},\ and\ \citenamefont {Collings}}]{horowitz2005}%
  \BibitemOpen
  \bibfield  {author} {\bibinfo {author} {\bibfnamefont {V.~R.}\ \bibnamefont
  {Horowitz}}, \bibinfo {author} {\bibfnamefont {L.~A.}\ \bibnamefont
  {Janowitz}}, \bibinfo {author} {\bibfnamefont {A.~L.}\ \bibnamefont {Modic}},
  \bibinfo {author} {\bibfnamefont {P.~A.}\ \bibnamefont {Heiney}}, \ and\
  \bibinfo {author} {\bibfnamefont {P.~J.}\ \bibnamefont {Collings}},\
  }\href@noop {} {\bibfield  {journal} {\bibinfo  {journal} {Physical Review
  E}\ }\textbf {\bibinfo {volume} {72}} (\bibinfo {year} {2005})}\BibitemShut
  {NoStop}%
\bibitem [{\citenamefont {Oswald}(2012{\natexlab{b}})}]{oswald7}%
  \BibitemOpen
  \bibfield  {author} {\bibinfo {author} {\bibfnamefont {P.}~\bibnamefont
  {Oswald}},\ }\href@noop {} {\bibfield  {journal} {\bibinfo  {journal}
  {Europhys. Lett.}\ }\textbf {\bibinfo {volume} {100}},\ \bibinfo {pages}
  {26001} (\bibinfo {year} {2012}{\natexlab{b}})}\BibitemShut {NoStop}%
\bibitem [{\citenamefont {Lavrentovich}\ and\ \citenamefont
  {Nastishin}(1987)}]{lavrentovich}%
  \BibitemOpen
  \bibfield  {author} {\bibinfo {author} {\bibfnamefont {O.~D.}\ \bibnamefont
  {Lavrentovich}}\ and\ \bibinfo {author} {\bibfnamefont {Y.~A.}\ \bibnamefont
  {Nastishin}},\ }\href@noop {} {\bibfield  {journal} {\bibinfo  {journal}
  {Ukr. Fiz. Zh.}\ }\textbf {\bibinfo {volume} {32}},\ \bibinfo {pages} {710}
  (\bibinfo {year} {1987})}\BibitemShut {NoStop}%
\bibitem [{\citenamefont {Akopyan}\ \emph {et~al.}(2001)\citenamefont
  {Akopyan}, \citenamefont {Alaverdian}, \citenamefont {Santrosian},\ and\
  \citenamefont {Chilingarian}}]{akopyan2}%
  \BibitemOpen
  \bibfield  {author} {\bibinfo {author} {\bibfnamefont {R.~S.}\ \bibnamefont
  {Akopyan}}, \bibinfo {author} {\bibfnamefont {R.~B.}\ \bibnamefont
  {Alaverdian}}, \bibinfo {author} {\bibfnamefont {E.~A.}\ \bibnamefont
  {Santrosian}}, \ and\ \bibinfo {author} {\bibfnamefont {Y.~S.}\ \bibnamefont
  {Chilingarian}},\ }\href@noop {} {\bibfield  {journal} {\bibinfo  {journal}
  {J. Appl. Phys.}\ }\textbf {\bibinfo {volume} {90}},\ \bibinfo {pages} {3371}
  (\bibinfo {year} {2001})}\BibitemShut {NoStop}%
\bibitem [{\citenamefont {Poursamad}(2009)}]{poursamad}%
  \BibitemOpen
  \bibfield  {author} {\bibinfo {author} {\bibfnamefont {J.~B.}\ \bibnamefont
  {Poursamad}},\ }\href@noop {} {\bibfield  {journal} {\bibinfo  {journal} {J.
  Contemp. Phys.}\ }\textbf {\bibinfo {volume} {44}},\ \bibinfo {pages} {14}
  (\bibinfo {year} {2009})}\BibitemShut {NoStop}%
\end{thebibliography}%

\end{document}